\documentclass[11pt]{article}
\usepackage{graphicx}
\usepackage{amssymb}
\usepackage{amscd}
\usepackage{mathrsfs}
\usepackage{longtable,lscape}
\usepackage{amsthm}
\usepackage{amsfonts}
\usepackage{amsmath}
\usepackage{bbm}
\usepackage{float}
\usepackage{url}

\newcommand{\captionfonts}{\footnotesize}
\makeatletter  % Allow the use of @ in command names
\long\def\@makecaption#1#2{%
  \vskip\abovecaptionskip
  \sbox\@tempboxa{{\captionfonts #1: #2}}%
  \ifdim \wd\@tempboxa >\hsize
    {\captionfonts #1: #2\par}
  \else
    \hbox to\hsize{\hfil\box\@tempboxa\hfil}%
  \fi
  \vskip\belowcaptionskip}
\makeatother 
\begin{document}
\title{Quantum cognition goes beyond-quantum: modeling the collective participant in psychological measurements}
\author{Diederik Aerts$^1$, Massimiliano Sassoli de Bianchi$^1$, Sandro Sozzo$^2$  and Tomas Veloz$^{1,3}$ \vspace{0.5 cm} \\ 
        \normalsize\itshape
        $^1$ Center Leo Apostel for Interdisciplinary Studies, 
         Brussels Free University \\ 
        \normalsize\itshape
         Krijgskundestraat 33, 1160 Brussels, Belgium \\
        \normalsize
        E-Mails: \url{diraerts@vub.ac.be}, \url{msassoli@vub.ac.be}
          \vspace{0.5 cm} \\ 
         \normalsize\itshape
        $^2$ School of Management and IQSCS, University of Leicester \\ 
        \normalsize\itshape
         University Road, LE1 7RH Leicester, United Kingdom \\
        \normalsize
        E-Mail: \url{ss831@le.ac.uk} 
        \vspace{0.5 cm} \\ 
        \normalsize\itshape
        $^3$ Universidad Diego Portales, Vicerrector\'ia Acad\'emica, Manuel Rodrí\'iguez Sur 415, \\ 
        \normalsize\itshape
        8370179 Santiago, Chile and Instituto de Filosof\'ia y Ciencias de la Complejidad, \\
	\normalsize\itshape
        \normalsize
	Los Alerces 3024, \~Nu\~noa, Santiago, Chile \\
        \normalsize
        E-Mail: \url{tveloz@gmail.com}
              }
\date{}
\maketitle
\begin{abstract}
\noindent In psychological measurements, two levels should be distinguished: the {\it individual level}, relative to the different participants in a given cognitive situation, and the {\it collective level}, relative to the overall statistics of their outcomes, which we propose to associate with a notion of {\it collective participant}.  When the distinction between these two levels is properly formalized, it reveals why the modeling of the collective participant generally requires beyond-quantum  -- \emph{non-Bornian}  -- probabilistic models, when sequential measurements at the individual level are considered, and this though a pure quantum description remains valid for single measurement situations. 
\end{abstract}

\section{Introduction\label{intro}}

The aim of mathematical psychology is to develop theoretical models of cognitive (and more generally psychological) processes~\cite{Coombs1970}. Its methodology comprises two main aspects. The first one is to establish sensible assumptions about how humans behave in the psychological/conceptual  
situations under study, which are then translated into specific mathematical descriptions. The second aspect is about confronting human subjects to these conceptual situations in controlled experimental settings, usually consisting of preliminary sessions where participants are instructed about what to do in the experiments, followed by the real situations where the data of their responses is collected, analyzed and then confronted with the models. Generally speaking, a mathematical model will be considered satisfactory if capable of explaining the collected data (like the relative frequencies of outcomes, interpreted as probabilities) and possibly predict their structure (the relations which data obey)~\cite{BusemeyerDiederich2009,Zucchini2000}. 

A paradigmatic case is the study of the \emph{conjunction fallacy}~\cite{Tversky1983}, where participants are observed to overall
violate the rules of classical probability theory, when asked to judge likelihood of events. Numerous competing mathematical models exist to explain the observed non-classicality of their estimations, including models based on Bayesian probability~\cite{Tentori2013}, various types of heuristics~\cite{Tversky1994} and more recently also quantum probability structures~\cite{Busemeyer2011}, and the success of these models is always contrasted in terms of how well they are able to account and explain the experimental probabilities. 

A subtle element that is usually overlooked, or not discussed, in the construction of models, is that in a cognitive psychology experiment participants may either all respond (actualize potential outcomes) according to a same mathematical model, or each one according to a different mathematical model. More precisely, in experiments participants may all actualize an outcome in the  \emph{same way} or in a  \emph{different way}. Also, depending on the situation they are 
confronted with, they can do this either in a \emph{deterministic way}, or in a \emph{indeterministic way}. By `deterministic' we mean a process whose outcome is in principle predictable in advance, whereas by `indeterministic' we mean a failure of predictability, i.e., a process where different possible outcomes are truly available, so that there is a situation of genuine unpredictability (the outcome cannot be determined in advance, not even in principle). When participants behave all deterministic, they can all choose the same outcome, or possibly different outcomes, and when they are all indeterministic, they can all actualize outcomes according to the same probabilistic model (like the one described by the quantum Born rule of probabilistic assignment), or according to different probabilistic models, describing more general (beyond-quantum, non-Bornian) ways to respond or take a decision (i.e., actualizing an outcome). Also, in the situation where they do not all behave in a statistically equivalent way, there can be a mixture of both deterministic and indeterministic ways of selecting an outcome. 

The main goal of the present article is to show that, when the above possibilities are considered, fundamentally different types of probabilistic models might be required to account for the aggregated experimental data. In other words, the modeling of the \emph{collective level} depends on the specificities of the \emph{individual level}, so that a comprehensive theory of psychological experiments/measurements must necessarily take into account the latter. This does not mean that one has to possess information about the individual processes of actualization of potential outcomes (the individual way of producing responses, or decisions), but the modeling of the situation at the individual level must be compatible with the modeling at the collective level. This highlights the importance of making a clear distinction between these two levels, the individual being associable with the participant's decisions, and the collective with a notion of \emph{collective participant}, 
whose behavior -- its way of actualizing outcomes -- is precisely described by the statistical indicators representing the entire collection of individual responses, in a unitary way.

In other words, in this article we pursue a new and definite approach to the foundations of cognitive psychology, and more specifically to the nature of psychological measurements, following our successful studies on the foundations of physical theories and measurement processes. By introducing the well-defined notion of collective participant, which was already implied in the {\it Brussels operational-realistic approach to cognition} (see \cite{AertsSassolideBianchiSozzo2015} and the references cited therein), but was never made fully explicit until now (and which we think is a new theoretical notion in the psychological landscape), we draw general and fundamental conclusions about the probabilistic structure characterizing cognitive phenomena. To this end, we start in the next section by listing the basic elements of typical psychological measurements, explaining how they can be properly formalized.

\section{The basic elements of psychological measurements \label{formalizing}}

We introduce the following basic elements characterizing psychological measurements:

{\it Number of participants}. A measurement always comprises a given number $n$ of participants (also called subjects, respondents, etc.) To obtain from them a significant statistics of outcomes, ideally $n$ will be a large number (in certain measurements it can just be several tens, in others it can reach a few thousands). 

{\it Uniform average}. The $n$ individuals participating in a measurement are all confronted with the same conceptual situation, from whom the experimenter obtains individual outcomes (e.g., the selection of items, yes-no answers, the assessment of rankings about given exemplars, etc.). These individual outcomes are then counted to calculate their relative frequencies, interpreted as probabilities (for $n$ sufficiently large), which are the main object of the mathematical modeling. 

{\it Individual level}. The description of the conceptual situations and changes happening at the level of each individual participating in the experiment. 

{\it Collective level}. The description of the conceptual situations and changes happening at the level where the different cognitive activities of the participants are considered as a whole, in particular when their outcomes are averaged out in the final statistics. 

{\it Intersubjectivity}. Each participant is submitted to the same experimental situation, which can be described as a \emph{conceptual entity $S$ prepared in a given state} \cite{AertsSassolideBianchiSozzo2015}. Such state is objective, or intersubjective, in the sense that each participant interacts with the same conceptual entity, prepared in the same initial state, describing the reality, or state of affair, of said entity.

{\it State-space representation}. For all participants, a same \emph{state-space} $\Sigma$ can be used to model the (conceptual entity describing the) cognitive situation. It can be a $\sigma$-algebra, a Hilbert space, a generalized Bloch sphere \cite{AertsSassolideBianchi2014a}, or any other suitable mathematical structure \cite{AertsSassolideBianchi2015a, AertsSassolideBianchi2015b, asdb2015f}. Consequently, the same element of $\Sigma$ is used to represent the state $p_{\rm in}$ of the conceptual entity $S$, describing the \emph{initial} conceptual situation to which all participants are equally submitted to.
 
{\it Outcomes}. A psychological measurement is characterized by a given number $N$ of possible outcomes, generally understood as the states to which the conceptual entity can transition to, when submitted to the action of a participant's mind subjected to an interrogation, or a decision-making process, specifying the outcomes (i.e., the answers, or the decisions) that are available to be actualized. Each participant is submitted to the same set of possible outcomes, as is clear that they are all part of the same measurement (or sub-measurement), and therefore have to follow the exact same protocol. So, in the same way a same state $p_{\rm in}$ is used to describe the initial condition of the measured conceptual entity $S$, also the outcome states will be represented by a same set $\{q_1,\dots, q_N\}$ of elements of $\Sigma$, for all participants. In the special case where the state space is a Hilbert space, this means that each participant is associated with the same spectral family.\footnote{This is clearly an idealization. But idealizations are what one tries to obtain when modeling experimental situations, as the scope of an idealization is precisely that of capturing the essential aspects of what a situation is all about, neglecting those aspects that are considered not to be essential. This is what physicists also do, for instance when modeling physical quantities (observables) by means of self-adjoint operators, or when describing state transitions by means of the L\"uders-von Neumann projection formula. Another useful idealization here, in analogy with so-called measurements of the first kind in physics, is to assume that measurement outcomes can be associated with well-defined outcome states. i.e., that after the measurement the measured entity is always left in a well-defined (eigen) state.}

{\it State change}. When submitted to a measurement (interrogative) context consisting of a conceptual entity $S$ in the initial state $p_{\rm in}$, with $N$ possible outcomes $q_j\in\Sigma$, $j=1,\dots,N$, each participant, by providing one (and only one) of these outcomes, changes the entity's initial state to the state 
corresponding to the selected outcome. In other words, a participant's cognitive action produces one of the following $N$ transitions: $p_{\rm in}\to q_j$, $j=1,\dots,N$. This action will be assumed to happen in a \emph{two-step process}, where the first step, if any, is deterministic, whereas the second step, if any, is indeterministic.

{\it Deterministic context}. Each participant, say the $i$-th one, starts her/his cognitive action by possibly changing the entity's initial state $p_{\rm in}$ in a deterministic way, producing a transition: $e^{\rm det}_i: p_{\rm in} \to p'_i$. One can for instance 
think of the deterministic context $e^{\rm det}_i$ as resulting from some kind of \emph{information supply}. It can be some externally obtained background information, directly supplied by the experimenter in the ambit of the experimental protocol, or some internal information retrieval, resulting from the participants' \emph{thinking activity}. The process could also correspond to the evocation of that portion of the persons' memory that need to be accessed in order to respond to the situation in question. In the Hilbert space language, $e^{\rm det}_i$ will be typically modeled 
by a unitary matrix, or by an orthogonal projection operator.

{\it Indeterministic context}. Following the possible information supply that produces the deterministic transition $p_{\rm in} \to p'_i$, and assuming that $p'_i$ is not already one of the outcome-states (i.e., an eigenstate of the measurement), the $i$-th participant will operate a genuinely indeterministic transition $e^{\rm ind}_i$: $p'_i \to p''_i$, where $p''_i\in \{q_1,\dots,q_N\}$ is one of the available outcome-states of the psychological measurement. One can think of the indeterministic context $e^{\rm ind}_i$ as being the result of some \emph{subconscious mental activity},\footnote{One should not deduce from this that the deterministic contexts $e^{\rm det}_i$ would therefore be necessarily associated, or exclusively associated, with conscious/controlled mental activities. In other words, our distinction between deterministic and indeterministic contexts is different from the distinction of so-called \emph{dual process theory}.} during which the $i$-th participant builds a mental condition of unstable equilibrium, resulting from the balancing of the different tensions between the state $p'_i$ of the conceptual entity and the available mutually excluding answers $q_j\in\Sigma$, $j=1,\dots,N$, competing with each other. This unstable equilibrium, by spontaneously breaking at some unpredictable moment, in an unpredictable way, then actualizes one of the possible outcome-states, in what can be described as a \emph{weighted symmetry breaking process} \cite{AertsSassolideBianchi2016}. 

Note that if there are no deterministic context effects, like (external or internal) information supply,  then $e^{\rm det}_i$ is just the trivial context, not affecting the initial state $p_{\rm in}$, so that $e^{\rm ind}_i$ will directly operate on the initial state $p_{\rm in}$. On the other hand, if the $i$-th participant knows in advance the answer to the interrogation, $e^{\rm ind}_i$ will be the trivial context, not affecting the state $p'_i$ obtained through the previous information retrieval activity (consisting in the subject looking into her/his memory, to discover the already existing answer to the question addressed, which s/he will simply select as the outcome). In general, we can think of $e^{\rm ind}_i$ as a context triggering a process that takes a very short time (something like a sudden collapse). On the other hand, the transition induced by $e^{\rm det}_i$ will generally require a longer amount of time to be produced, corresponding to the time the $i$-th participant needs to obtain and assimilate the background information, before selecting an outcome. So, ideally and generally speaking, we will assume 
that to each individual participating in a measurement we can attach a context $e_i$, which can be 
understood as the composition of two contexts, one deterministic and the other indeterministic: 
\begin{equation}
e_i=e^{\rm ind}_i\circ e^{\rm det}_i, \quad i\in\{1,\dots,n\}.
\label{decomposition}
\end{equation}

It is important to specify that although the above decomposition, and the previous specifications of deterministic and indeterministic contexts, are meant to express the relative generality of our approach, they should not be misunderstood as a claim that all measurement situations should or could be modeled in this way. That said, we think it may be useful to give a couple of examples. Consider a survey including the question: ``Are you a smoker or a non-smoker?''. Obviously, $e^{\rm ind}_i$ will be then the trivial contexts, as each participant knows in advance if s/he is a smoker or a non-smoker, so only non-trivial deterministic contexts $e^{\rm det}_i$ will be present, whose action will simply be the retrieval of the already existing answer about the participant's smoking or non-smoking habits. Consider now the question \cite{aa1995, Aertsetal1999}: ``Are you for or against the use of nuclear energy?'' In this case, non-trivial indeterministic contexts will be present, being very plausible that some of the participants have no prior opinion about nuclear energy, hence for them the outcome will be literally created (actualized) when answering the question, in a way that cannot be predicted in advance (as in a symmetry breaking process). Imagine also that before the question is asked some literature is given to the participants to read, explaining how greenhouse gas emissions can be reduced through the use of nuclear energy. This will obviously influence in a determined way how decisions will be taken at the individual level, increasing the percentage of those who will answer by favoring nuclear energy (with some of them even possibly becoming fully deterministic in the way they will answer). So, we are in a situation where both non-trivial deterministic and non-trivial indeterministic contexts are both possibly present and operate in a sequential way, in accordance with the decomposition (\ref{decomposition}).

\section{Ways of choosing and background information\label{ways}}

For the sake of clarity, in the following we shall simply call $e^{\rm det}_i$ the \emph{background information} and $e^{\rm ind}_i$ the \emph{way of choosing} of the $i$-th subject, respectively. Let then $\mu(q_j,e_i,p_{\rm in})\geq 0$ be the probability that the $i$-th participant produces the outcome $q_j$, via the individual context $e_i$, when submitted to the conceptual entity in the  initial state $p_{\rm in}$. We clearly have: $\sum_{j=1}^N \mu(q_j,e_i,p_{\rm in}) =1$, for 
every individual context $e_i$ and every  initial state $p_{\rm in}\in \Sigma$.

It is at this stage useful to distinguish the following three situations.

{\it Situation 1: The way of choosing is trivial.} This corresponds to the situation where all participants choose in a (in principle) predictable way, i.e., $e_i=e^{\rm det}_i$, for every $i$, which however can be different for each participant. This means that the values of the probabilities ${\mu}(q_j,e_i,p_{\rm in})$ are all either $0$ or $1$. However, if the number of individuals deterministically selecting the outcome $q_j$ is $n_j$, then at the collective level the experimental probability $\mu(q_j,e,p_{\rm in})$ for the transition $p_{\rm in}\to q_j$ is simply given by the ratio ${n_j\over n}$, where $e$ denotes the context of the \emph{collective participant} 
associated with the collection of $n$ individual participants. Here we can distinguish two cases: (a) $e^{\rm det}_i = e^{\rm det}$, for every $i=1,\dots,n$; (b) for some $i$ and $j$, we can have $e^{\rm det}_i \neq e^{\rm det}_j$. In case (a), the process is deterministic also at the collective level. In case (b), since some of the probabilities $\mu(q_j,e,p_{\rm in})$ are different from $0$ or $1$, the process becomes indeterministic at the collective level.

{\it Situation 2: The background information is trivial.} This corresponds to the situation where there is no information supply (or other deterministic context effects) 
prior to the selection of an outcome. We then have $e_i=e^{\rm ind}_i$, for every $i=1,\dots,n$, so that the different participants only produce an indeterministic (quantum-like) transition. Again, we can distinguish two cases: (a) participants all choose in the same way, which of course does not mean they all choose the same outcome, i.e., $e^{\rm ind}_i = e^{\rm ind}$, for every $i=1,\dots,n$; (b) participants possibly choose in different ways, i.e., for some $i$ and $j$, we can have $e^{\rm ind}_i \neq e^{\rm ind}_j$.

{\it Situation 3: The way of choosing and the background information are both non-trivial.} This is the most general and complex situation, and the following two cases can be distinguished: (a) $e^{\rm det}_i = e^{\rm det}$ and $e^{\rm ind}_i = e^{\rm ind}$, so that $e_i = e$, for every $i=1,\dots,n$, with $e=e^{\rm ind}\circ e^{\rm det}$; (b) for some $i$ and $j$, we can have $e^{\rm det}_i \neq e^{\rm det}_j$ and/or $e^{\rm ind}_i \neq e^{\rm ind}_j$. 

Coming back to the examples given at the end of the previous section, Situation 1 corresponds to participants being for instance asked if they are smokers. If they are all, then we are in case (a), otherwise, if there is a mix of smokers and non-smokers, we are in case (b). Situation 2 corresponds to participants being for instance asked if they are in favor of nuclear energy, assuming that none of them ever reflected or took a position on the matter before having to answer such question. If they are a very homogeneous group, say of same sex, cultural background, age group, etc., then we can assume that (at least in first approximation) we are in case (a), whereas if there are relevant differences among the participants, for instance because a portion of them do not even know what nuclear energy is, then we are in principle in case (b). Situation 3 corresponds to participants being for instance asked to answer a preliminary question (before being asked the nuclear energy question), to which all of them can respond in a predictable way. For example, the question could be: ``Have you ever heard of Chernobyl's disaster?'' Clearly, either participants have heard about it, or not. If they are a homogeneous group, we can assume they will all answer in the same way, say in an affirmative way, and that they will subsequently answer the nuclear energy question with same individual probabilities (which will have been altered in a deterministic way by the previous question, for instance because it contains the word ``disaster''). So, this would be an example of case (a) in Situation 3, whereas case (b) would be when either participants are asked different preliminary questions, and/or when the group is non-homogeneous.

A few remarks are in order. Note that for all three situations above, in case (a) there is no difference between the individual and collective level, i.e., all participants are ``cognitive clones,'' behaving exactly in the same way, whereas in the general case (b), they can produce distinct individual behaviors. 

Different individual background information can be easily modeled in a Hilbert space representation. Indeed, it is sufficient to associate to each participant (or group of participants) a different unitary evolution \cite{asdb2017b}, or a different projection operator \cite{a2007}. On the other hand, different individual ways of choosing cannot be modeled by remaining within the confines of standard quantum mechanics, as the latter only admits the \emph{Born rule} way of choosing, apparently imposed by Gleason's theorem. We say ``apparently" because Gleason's theorem only tells us that if the transition probabilities only have to depend on the state before the measurement and on the eigenstate actualized after the measurement, then they must be given by the Born rule. However, if we relax this constraint, one can introduce different parameter-dependent probability measures within a same Hilbert (state) space representation, and use them to describe the different individual ways of choosing of the participants. In other words, by extending the standard Hilbertian formalism (see for instance \cite{AertsSassolideBianchi2014a, AertsSassolideBianchi2015a, AertsSassolideBianchi2015b, asdb2017a}), it becomes possible, while maintaining the Hilbertian structure for the states, to define different rules of probabilistic assignment, characterizing the participants' different ways of choosing (see also Sec.~\ref{two-outcome}).

One may wonder what could be the meaning of an individual statistics of outcomes. We know that from the outcomes provided by all the participants we can determine the experimental probabilities, by calculating their relative frequencies. This is the description of what we have called the collective level, associated with the notion of collective participant. 
However, if we assume that the individual contexts $e_i$ are generally non-deterministic, it is natural to also associate outcome probabilities ${\mu}(q_j,e_i,p_{\rm in})$, $j=1,\dots N$, $i=1,\dots,n$, to each one of the $n$ individuals participating in the experiment. Of course, this does not mean that these probabilities can be directly or easily determined. However, we also know that individuals have the ability to provide not only a specific answer to a given question, but also to estimate the probabilities for the available answers, which is a strong indication that it is correct to associate each individual with a specific statistics of outcomes. So, when we calculate the relative frequencies ${n_j\over n}$ of the different outcomes, what we are in fact estimating is the probabilistic average: 
\begin{equation}
\mu(q_j,e,p_{\rm in})={1\over n}\sum_{i=1}^n \mu(q_j,e_i,p_{\rm in}), \quad j=1,\dots N,
\label{average}
\end{equation}
where $e$, as we already mentioned above, denotes the context of the \emph{collective participant}.

\section{The collective participant\label{meta}}

At this stage of our analysis, it is important to emphasize a difference between the present operational-realistic description of psychological measurements and the today most commonly adopted (subjectivist) view in quantum cognition (and cognitive modeling in general), according to which the initial state would describe the \emph{belief system} of a participant about the cognitive situation under consideration. Here a question arises: Which participant? Consider for a moment the typical situation of a quantum measurement in a physics laboratory: a same measurement apparatus (and usually a same agent operating on it) is used $n$ times, with the physical entity always prepared in the same state, to obtain the final statistics of outcomes. On the other hand, in a psychological laboratory $n$ different participants play the role of the apparatus, i.e., we actually have $n$ different measurement apparatuses used in a same experiment, and each one is typically used only once. So, if we want psychological measurements to be interpretable in a manner analogous to physics measurements, the $n$ participants must be considered to be like clones, i.e., like measuring entities all having the same way of accessing the available background information, and the same way of choosing an outcome, the latter being described for instance by the Born rule. Also, one must assume that they all have the same way of updating their probabilities for subsequent measurements, by associating to their answers the same outcome states. 

A first difficulty is that there are no reasons to think that all participants will necessarily share the exact same belief regarding the cognitive situation they are subjected to, i.e., that they will represent such situation by using the same vector in Hilbert space, and the same is true regarding the choice of the outcome states. On the other hand, as we mentioned already, in our approach the state describes an aspect of the reality of the conceptual entity under consideration, in a given moment, i.e., an intersubjective element of the cognitive domain shared by all the participants \cite{AertsSassolideBianchiSozzo2015}. Therefore, the above difficulty does not arise in a realistic approach. However, the fact remains that a psychological measurement is performed by $n$ participants, with different mindsets, and not just by one participant, i.e., by a single mind. Each of them will behave in a different way when confronted with the cognitive situation, i.e., will elicit one of the outcomes by pondering and choosing in possibly different ways. In other words, if the $n$ participants are not assumed to behave as equivalent measurement apparatuses, i.e., are not assumed to be statistically equivalents, then we certainly cannot consider the experimental probabilities to be descriptive of their individual actions (or, in the subjectivist view, of their individual beliefs and judgments). 

What we mean to say is that the states and probabilities describing the overall statistics generated by the collection of $n$ individuals can only be associated with a \emph{memoryless collective participant},  such that if it would be submitted $n$ times to the interrogative context in question, the statistics of outcomes it would produce would be equivalent (for $n$ large enough) to the statistics of outcomes generated by the $n$ participants in the experiment. The collective mind of such collective participant has therefore to be understood as a composite entity formed by $n$ separate sub-minds. When interrogated, it provides an answer by operating in the following way: first, it selects one of its internal sub-minds, say the $i$-th one (which works exactly as the mind of the $i$-th participant), then, it uses it to answer the question and to produce an outcome, and if asked again the question, it selects another of its internal sub-minds, among the $n-1$ that haven't yet provided an answer, and so on. This means that, for as long as the same question is asked no more than $n$ times, the mind of the collective participant  
will show no memory effects (for instance, no response replicability effects). This memoryless property of the collective participant, 
associated with the overall statistics of outcomes, is what fundamentally distinguishes it from the individual participants. But apart from that, in many situations one can certainly describe the cognitive action of the collective participant in a way that is analogous to the description of the individual ones.

Now, when referring to the great success of quantum cognition, one usually points to the success of the standard quantum formalism in modeling the cognitive action of the collective participant. So, a natural question arises: Why this success? Many answers can of course be given (see for instance the first chapter of \cite{bb2012}). Let us briefly explain the answer to which we arrived in \cite{AertsSassolideBianchi2015a,AertsSassolideBianchi2015b}. For this, we consider \emph{Situation 2} of Sec.~\ref{ways}, where the participants select an answer without acquiring any prior information. Then, the question is: Considering that the individual contexts $e_i$ (here equal to $e^{\rm ind}_i$) are in principle all different, and therefore not describable by means of the Born rule, why is it nevertheless possible, in general, to describe the context $e$, associated with the collective participant,  by means of the latter? In other words: Why the averages (\ref{average}) are usually well described by the quantum mechanical transition probabilities? 

To answer this question, one has first to find a way to describe all possible ways of choosing an outcome. This can be done by exploiting the \emph{general tension-reduction} (GTR) model \cite{AertsSassolideBianchi2015a,AertsSassolideBianchi2015b,asdb2015f}, or its more specific implementation called the \emph{extended Bloch representation} (EBR) of quantum mechanics \cite{AertsSassolideBianchi2014a}, where the set of states is taken to be Hilbertian. Then, one has to find a way to calculate the average over all possible ways of choosing an outcome (called a \emph{universal average}). This can be done by following a strategy similar to that used in the definition of the \emph{Wiener measure}, and the remarkable result is that the average probabilities so obtained are precisely those predicted by the Born rule (if the state space is Hilbertian), thus explaining why the latter generally appears as an optimal approximation in numerous experimental situations \cite{AertsSassolideBianchi2014a,AertsSassolideBianchi2015b}. In other words, as $n$ increases, the average (\ref{average}) can be expected to become better and better approximated by the Born rule, i.e., the context $e$, characterizing the action of the collective participant,  
is expected to tend towards that context that is described by the Born rule (and the associated projection postulate).

\section{Two-outcome measurements\label{two-outcome}}

\begin{figure}[htbp]
\begin{center}
\includegraphics[width=15cm]{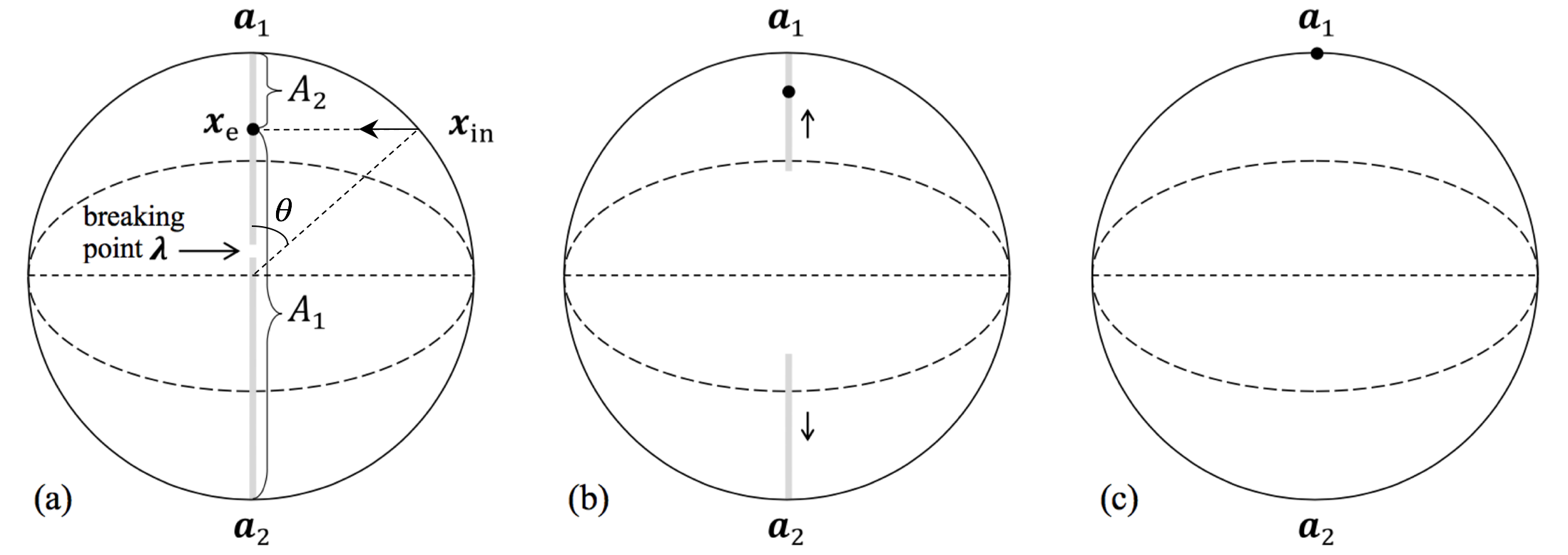}
\caption{A schematic description of the two phases of a two-outcome measurement, in the EPR of quantum mechanics. Here the breaking of the abstract elastic structure happens in $A_1$, so that the outcome is ${\bf a}_{1}$.} 
\label{Figure1}
\end{center}
\end{figure}

Let us more specifically explain, in the simple situation of two-outcome processes, how different kinds of measurements, associated with a same initial state and 
pair of outcomes, can be modeled within the EBR. We consider the 3-dimensional Bloch sphere representation of states, with the initial state $p_{\rm in}$ described by a unit 3-dimensional real vector ${\bf x}_{\rm in}$. To fix ideas, one can imagine a virtual point-particle associated with it, i.e., positioned exactly at point ${\bf x}_{\rm in}$ on the surface of the unit sphere, which should also be imagined as an hollow structure in which the particle can penetrate. A measurement having two outcome-states $q_1$ and $q_2$ can then be represented as a one-dimensional simplex $\triangle_1$ with apex-points ${\bf a}_{1}$ and ${\bf a}_{2}$, corresponding to the two Bloch vectors representative of states $q_1$ and $q_2$, respectively 
(see Figure~\ref{Figure1}). One can think of $\triangle_1$ as an abstract elastic and breakable structure, extended between the two points ${\bf a}_{1}$ and ${\bf a}_{2}$. A measurement process can then be described as a two-phase process. During the first phase, the point particle enters the sphere, following a path orthogonal to $\triangle_1$, thus reaching a point ${\bf x}_{\rm e}=({\bf x}_{\rm in}\cdot {\bf a}_{1})\, {\bf a}_{1}$. Cognitively speaking, one 
can think of this first phase as a process during which the mind of an individual brings the situation described by $p_{\rm in}$ into the context of the two possible answers $q_1$ and $q_2$, i.e., as a (deterministic) preparation process during which it brings the meaning of the situation as close as possible to the meaning of the possible answers. 

The second phase consists in the elastic breaking at some unpredictable point \mbox{\boldmath$\lambda$}, so that its subsequent collapse  can 
bring the abstract point particle either towards point ${\bf a}_{1}$ or point ${\bf a}_{2}$, depending on whether \mbox{\boldmath$\lambda$} belongs to the segment $A_1$, between ${\bf x}_{\rm e}$ and ${\bf a}_{2}$, or to the segment $A_2$, between ${\bf a}_{1}$ and ${\bf x}_{\rm e}$, respectively. Cognitively speaking, this second (typically indeterministic) phase corresponds to the reduction of the tensional equilibrium previously built, due to fluctuations causing the breaking of such equilibrium and consequent selection of only one of the two possible answers. 

Being $\triangle_1$ of length $2$, we can parametrize its points using the interval $[-1,1]$, where the values $x=-1$ and $x=1$ correspond to vectors ${\bf a}_{2}$ and ${\bf a}_{1}$, respectively, and $x_{\rm e}= {\bf x}_{\rm in}\cdot {\bf a}_{1}=\cos\theta$ to the position ${\bf x}_{\rm e}$ of the particle once it has reached the elastic, so that $A_1=[-1,\cos\theta]$ and $A_2=[\cos\theta,1]$ (see Figure~\ref{Figure1}). If $\rho(y)$ is the probability density describing the way the elastic can break, then the probability for the transition $p_{\rm in}\to q_1$ (i.e., the probability for the abstract point particle to go from the initial position ${\bf x}_{\rm in}$ to the final position ${\bf a}_{1}$, passing through the equilibrium point ${\bf x}_{\rm e}$) is given by: 
\begin{equation}
\label{rhoprobabilitymeasurement}
\mu(q_1,e_\rho,p_{\rm in})=\int_{-1}^{\cos\theta} \rho(y)dy,
\end{equation}
where $e_\rho$ is the context associated with the $\rho$-way of breaking of the elastic, and of course $\mu(q_2,e_\rho,p_{\rm in})=1-\mu(q_1,e_\rho,p_{\rm in})$. 

It is worth observing that, depending on the breakability of the elastic, $e_\rho$ will be either a deterministic or indeterministic context \cite{Aertsetal1997,Aerts1998,Aertsetal1999}. For instance, if the elastic can only break in the segment $[-1,x]$, with $x < \cos\theta$, it immediately follows from (\ref{rhoprobabilitymeasurement}) that $\mu(q_1,e_\rho,p_{\rm in})=1$, so we are in the situation of a deterministic context. Note that such situation cannot be described by standard quantum mechanics, since a probability equal to 1, in a first kind measurement, is only possible if the initial state is an eigenstate, which is not necessarily the case here. Of course, if the elastic can instead only break in $[x,1]$, with $x > \cos\theta$, then $\mu(q_1,e_\rho,p_{\rm in})=0$, so $e_\rho$ still describes a deterministic process, but this time with the predetermined outcome being $q_2$. On the other hand, if the elastic has breakable parts both in $A_1$ and $A_2$, the outcome cannot be predicted in advance and $e_\rho$ describes a genuine indeterministic context. Note that a context $e_\rho$ can be deterministic for a given state $p_{\rm in}$, associated with a given angle $\theta$, and indeterministic for a different state $p'_{\rm in}$, associated with a different angle $\theta'$. Note  that the above sphere-model can also describe measurements that are deterministic for all initial states, if the elastic that can only break in one of its two anchor points, or indeterministic for all initial states, if the elastic that can only break, with some given probabilities, in the two anchor points \cite{Aerts1998, AertsSassolideBianchi2015a}.

We can now state in more precise terms the content of the above mentioned result: when taking the universal average of (\ref{rhoprobabilitymeasurement}), i.e., when averaging over all possible $\rho$-ways an elastic can break -- let us denote such averaged probability $\langle\mu(q_1,e_\rho,p_{\rm in})\rangle^{\rm univ}$ -- one can show that the latter is identical to the probability associated with a uniformly breaking elastic, characterized by the constant probability density $\rho_{\rm u}(y)={1\over 2}$. In other words \cite{AertsSassolideBianchi2014a,AertsSassolideBianchi2015b}: 
\begin{equation}
\label{average-univ}
\langle\mu(q_1,e_\rho,p_{\rm in})\rangle^{\rm univ}=\mu(q_1,e_{\rho_{\rm u}},p_{\rm in})={1\over 2}\int_{-1}^{\cos\theta} dy= {1\over 2}(1+\cos\theta).
\end{equation}
We observe that the universal average (\ref{average-univ}) is identical to the Born quantum probability.\footnote{The average (\ref{average-univ}) can also be interpreted as the outcome probability of a so-called \emph{universal measurement}, characterizing by a two-level ``actualization of potential properties'' process, i.e., such that there is not only the actualization of a measurement-interaction (the breaking point \mbox{\boldmath$\lambda$}), but also of a way to actualize a measurement-interaction (the probabilty density $\rho$). At the present state of our knowledge, it is an open question to know if quantum measurements are universal measurements.} In other words, when preforming a universal average, one recovers the quantum mechanical \emph{Born rule}, if the state space has an Hilbertian structure (i.e., is a Blochean representation derived from the Hilbert space geometry), 
which in part explain why Hilbert-models based on the Born rule can be used to efficaciously account for many experimental situations, also beyond the domain of microphysics \cite{AertsSassolideBianchi2015a,AertsSozzo2012a,AertsSozzo2012b}. 

So, if the average (\ref{average}) is performed on a sufficiently large sample of persons, each one describing a different way of selecting an outcome, one 
can expect $\mu(q_j,e,p_{\rm in})\approx \langle\mu(q_2,e_\rho,p_{\rm in})\rangle^{\rm univ}$, i.e., one can expect the collective participant  
to behave as a \emph{universal participant}, described by the Born probabilities. 

The above, however, can only work for as long as the averages are performed on single measurements. Indeed, if sequential measurements are considered at the level of the individual participants, then the situation becomes much more complex (the average defining the collective participant's 
probabilities becomes much more involved) and one cannot expect anymore the standard quantum formalism to be able to model all possible experimental situations, as we will show by means of a simple example in Section~\ref{violating-quantum-equalities}. But before that, let us describe the experimental situations where sequential measurements are performed (Sec.~\ref{sequential}), and what are the different modeling options (Sec.~\ref{modeling-virtual}). We will then derive a well-known quantum equality (Sec.~\ref{quantum-test}) and show that it can be easily violated if the individual participants are not all ``quantum clones.''

\section{Sequential measurements\label{sequential}}

We consider two psychological measurements, which we denote $A$ and $B$, and we assume that the $M_A$ outcome states of $A$ are $\{a_1, \dots, a_{M_A}\}$, whereas the $M_B$ outcome states for $B$ are $\{b_1, \dots, b_{M_B}\}$. In many experimental situations, it is observed that the outcomes probabilities obtained when $A$ is performed first, and then $B$, are not the same as those obtained when $B$ is performed first, and then $A$. More precisely, \emph{question order effects} manifest in the fact that, in general, the probability for the sequential outcome `$a_j$ then $b_k$' (denoted $a_jb_k$ in the following), when the sequence of measurements `$A$ then $B$' (denoted $AB$) is considered, is different from the probability for the sequential outcome `$b_k$ then $a_j$' (denoted $b_ka_j$ in the following), when the sequence `$B$ then $A$' is considered (denoted $BA$) \cite{asdb2017a, plosone,bb2012,WangBusemeyer2013}. Clearly, the collective participant  
to whom the sequential outcome probabilities are to be associated with, cannot be the one we previously described, as its main characteristic was that of being memoryless, whereas to produce order effects some short-term memory is needed. So, the situation is now more complex, which is the reason why, as we are going to see, a non-Hilbertian probability structure will generally emerge. 

In the practice, sequential measurements are executed on a same participant. For instance, assuming that we have $2n$ participants, half of them will be subjected to the sequence $AB$ and the other half to the sequence $BA$. More precisely, if $i\in\{1,\dots n\}$, then the $i$-th individual will first be submitted to measurement $A$ and then, immediately after, to measurement $B$ (which for instance are two sequential questions in an opinion poll). The outcome of the first measurement (the answer to the first question) can change the initial state of the conceptual entity for the second one, as the two measurements are assumed to be performed in a rapid succession (i.e., the two questions are asked one after the other), so that the outcome of the first will remain in the sphere of consciousness of the $i$-th participant when submitted to the second measurement (i.e., when answering the second question). Similarly, if $i\in\{n+1,\dots 2n\}$, then the $i$-th participant will first be submitted to measurement/question $B$ followed by measurement/question $A$.  If $n(a_jb_k)$ and $n(b_ka_j)$ are the total counts for the sequential outcomes $a_jb_k$ and $b_ka_j$, respectively, we have the experimental probabilities (relative frequencies): 
\begin{equation}
p(a_jb_k) ={n(a_jb_k)\over n},\quad p(b_ka_j) ={n(b_ka_j)\over n},
\label{exp-prob}
\end{equation} 
which in general will exhibit order effects, i.e., $p(a_jb_k)\neq p(b_ka_j)$, for some $j$ and $k$. These are the probabilities one will typically attempt to model using the quantum formalism. 

Since the sequential measurements are performed at the level of the individual participants, we can generally write: 
\begin{equation}
p(a_jb_k)= {1\over n}\sum_{i=1}^n \mu(a_j,e_{A,i},p_{\rm in})\mu(b_k,e_{B,i},a_j),\quad p(b_ka_j)= {1\over n}\sum_{i=n+1}^{2n} \mu(b_k,e_{B,i},p_{\rm in})\mu(a_j,e_{A,i},b_k),
\label{average2}
\end{equation}
where $e_{B,i}$ denotes the context associated with the $i$-th individual, when submitted to the $B$-measurement, which needs not to be equal to the context $e_{A,i}$ responsible for the outcomes of the $A$-measurement. Now, it is clear that the averages (\ref{average2}) are very different from (\ref{average}), as they do not consist in a sum of probabilities (characterizing the different individual contexts) all associated with the same initial and final state transition. Here we have a much more intricate sum of products of probabilities, associated with different state transitions, where the different contexts and states get mixed in a complicate way. This is a very different situation than that of a single measurement 
universal average, and one cannot expect anymore the overall statistics of outcomes to be well approximated by the Born rule, even when $n$ becomes very 
large. 

Consider once more a collective participant  
to be associated with the experimental probabilities (\ref{exp-prob}), i.e., such that, if submitted to the sequential measurements in question, it would deliver those same probabilities. We can write: 
\begin{equation}
p(a_jb_k)= \mu(a_j,e_A,p_{\rm in})\mu(b_k,e_B,a_j),\quad p(b_ka_j)= \mu(b_k,e_B,p_{\rm in})\mu(a_j,e_A,b_k).
\label{averages-virtual}
\end{equation} 
This time, however, it is not the collective mind of the collective participant  
that is subjected to the sequential measurements. Indeed, if this would be the case, no relevant order effects would be observed, as the sub-minds selected to answer the first question would not be the same as those answering the second question. So, the collective participant, in this case, turns out to be a much more artificial construct, as it would operate differently than how it does when subjected to a single measurement situation. Indeed, in the latter case each repetition of the measurement is performed by a different sub-mind (a different individual participant), whereas in the case of sequential measurements the same sub-mind is used to answer the two questions in the sequence.\footnote{One could imagine here that the sequential measurements are performed so swiftly one after the other that the collective participant  does not have the time to activate a different sub-mind to answer the two questions in the sequence, whereas enough time would be available in-between the sequences.} Also, as we said already, even if $n$ is large, there are no reasons to expect that the two contexts $e_A$ and $e_B$, characterizing the collective participant in relation to the $A$ and $B$ measurements, would be both describable by the Born rule, hence be also equal. In fact, because of the mixing between states and contexts in the sums (\ref{average2}), we cannot even expect in this case the collective participant to use the same vectors as the individual participants to describe the initial and final states, in the sense that, to be able to find two contexts $e_A$ and $e_B$ modeling the experimental probabilities as per (\ref{averages-virtual}), one will generally need to use an effective description for the states that is different from the one that is inter-subjectively employed by the individual participants (see the penultimate section of \cite{asdb2017a} for a discussion of this point).

\section{Different modeling options\label{modeling-virtual}}

Let us more specifically consider \emph{Situation 3} of Sec.~\ref{ways}, where participants not only can take decisions in different ways, but also be supplied with 
some information before doing so. We emphasize again that if one introduces deterministic processes that can change the initial state into a state which is possibly different for each participant, these have to be associated with clearly identifiable processes, which should be in principle predictable in advance (for instance, having access to sufficient information about the cultural background and  psychological profile of each individual). This because the contexts $e^{\rm det}_i$ are defined to be deterministic not only for the reason that they change the initial state in a predetermined way, but also because they are given in advance. If this would not be the case, i.e., if they would be actualized at the moment, when the participants are submitted to the conceptual situation, then they would be fundamentally indeterministic and their description should be included in $e^{\rm ind}_i$.

What about the collective participant in this case? Should we also associate it with a process of information supply, i.e., with a deterministic context $e^{\rm det}$, if this happens at the level of the individual participants? To answer this question, we assume for simplicity that there are only two intakes of information during the measurement, and more precisely that $n_a$ among the $n$ participants receive some information characterized by the context $e^{\rm det}_a$, whereas the remaining $n_b=n-n_a$ receive some information characterized by the context $e^{\rm det}_b$. Then, if $p_a$ is the state obtained when the initial state $p_{\rm in}$ is submitted to context $e^{\rm det}_a$, and we simply write $p_a= e^{\rm det}_a p_{\rm in}$, and similarly $p_b =e^{\rm det}_b p_{\rm in}$ is the state obtained when the initial state $p_{\rm in}$ is submitted to context $e^{\rm det}_b$, (\ref{average}) becomes:

\begin{equation}
\mu(q_j,e,p_{\rm in})={1\over n}\sum_{i=1}^{n_a} \mu(q_j,e^{\rm ind}_i,p_a) + {1\over n}\sum_{i=n_a+1}^{n} \mu(q_j,e^{\rm ind}_i, p_b).
\label{average-sum}
\end{equation}
To further simplify the discussion, we also assume that all participants choose in a trivial way (\emph{Situation 1} of Sec.~\ref{ways}) and that, say, $p_a=q_1$ and $p_b=q_2$. Then, $\mu(q_1,e,p_{\rm in})={n_a\over n}$ and $\mu(q_2,e,p_{\rm in})={n_b\over n}$. This means that, as we observed already, different deterministic cognitive actions performed by the individual participants translate, at the level of the collective participant, in an indeterministic action, to be described as a `way of choosing' and not as an `information supply'. Of course, the situation is different if all participants would access exactly the same information, as in this case one can also do as if the same would happen at the level of the collective participant. 

In the situation of sequential measurements, however, it is much less clear if the information supply process should also be included in the modeling of the collective participant, even though all individual (real) participants access the same information. Assume for instance that we are in the situation where before measurement $A$ some preliminary background information is given, which is the same for all participants, changing the initial state according to the deterministic context $e^{\rm det}_A$, and that following measurement $A$, and before measurement $B$, some further information is given, changing the outcome state of measurement $A$ according to the deterministic context $e^{\rm det}_B$, and same thing when the order of the measurements is changed (an example of this kind of situation is the so-called \emph{Rose/Jackson experiment} \cite{asdb2017a,WangBusemeyer2013,Moore2002,WangEtal2014}). Then we can write: 
\begin{eqnarray}
p(a_jb_k)= {1\over n}\sum_{i=1}^n \mu(a_j,e^{\rm ind}_{A,i},e^{\rm det}_A p_{\rm in})\mu(b_k,e^{\rm ind}_{B,i},e^{\rm det}_B a_j),
\nonumber\\
p(b_ka_j)= {1\over n}\sum_{i=1}^n \mu(b_k,e^{\rm ind}_{B,i},e^{\rm det}_B p_{\rm in})\mu(a_j,e^{\rm ind}_{A,i},e^{\rm det}_A b_k),
\label{average4}
\end{eqnarray}
and the question is: At the level of the collective participant, 
should we model the probabilities (\ref{average4}) by writing: 
\begin{equation}
p(a_jb_k)= \mu(a_j,e^{\rm ind}_A,e^{\rm det}_A p_{\rm in})\mu(b_k,e^{\rm ind}_B,e^{\rm det}_B a_j),\quad
p(b_ka_j)= \mu(b_k,e^{\rm ind}_B,e^{\rm det}_B p_{\rm in})\mu(a_j,e^{\rm ind}_A,e^{\rm det}_A b_k),
\label{option1}
\end{equation}
i.e., by also associating the collective participant  
with information supply deterministic contexts, or should we instead write:
\begin{equation}
p(a_jb_k)= \mu(a_j,e^{\rm ind}_A, p_{\rm in})\mu(b_k,e^{\rm ind}_B,a_j),\quad
p(b_ka_j)= \mu(b_k,e^{\rm ind}_B, p_{\rm in})\mu(a_j,e^{\rm ind}_A, b_k),
\label{option2}
\end{equation}
without including deterministic contexts at the collective level?

The modeling option (\ref{option1}) can be defended by saying that since the same information is given to all participants, and that the collective participant  
is meant to represent their overall behavior, the description of its virtual cognitive action should also include the deterministic contexts $e^{\rm det}_A$ and $e^{\rm det}_B$, in an explicit way. On the other hand, the modeling option (\ref{option2}) can be defended by saying that we can consider the information accessed by the participants before answering the questions to be part of the questions themselves. Indeed, questions always have some built-in context, i.e., some background information, appearing in more or less explicit terms in the way they are formulated. So, these deterministic contexts should be integrated in the indeterministic ones (see \cite{asdb2017a} for a discussion of this point in the ambit of the Rose/Jackson measurement). 

Another possible argument in favor of the modeling option (\ref{option2}) is the following. Since the averages (\ref{average4}) result from a sum of products of probabilities, so that the effects of the deterministic and indeterministic contexts are mixed in a complicate way, it is questionable if one should really attribute to the collective participant the same `information supply' processes of the individual participants (also considering that the former will not generally use the same representation for the initial and outcome states). In fact, one can go even further and question if it is really meaningful to model the experimental probabilities $p(a_jb_k)$ and $p(b_ka_j)$ as the products (\ref{option1}) or (\ref{option2}). Indeed, being the same individual who answers the sequence of questions, one could object that the correct way to interpret the experimental situation is to say that the $AB$ and $BA$ measurements are in fact single measurements with $M_AM_B$ outcomes each. This because the fact that a same participant (a same sub-mind of the collective participant) 
answers both questions should maybe be considered more relevant than the fact that they provide the answers in a sequential way. 

According to this last viewpoint, it would not be (or only be) the sequentiality of the answers at the origin of the observed order effects, in the sense that we can imagine a slightly different experimental context where each participant would be jointly submitted to both questions, and jointly provide a couple of answers, and it is not unreasonable to expect that it is the very fact that in the $AB$ and $BA$ measurements the couples of questions are presented in a different order that would be at the origin of the difference (or part of the difference) between the probabilities for the outcomes $a_jb_k$ and $b_ka_j$. In other words, the order effects would originate (or in part originate) at the level of how questions are formulated, as is clear that the order of the different statements contained in a sentence can be relevant for what concerns its perceived meaning. Take the following example \cite{Rampin2005}: 

\begin{quote}
A novice asked the prior: ``Father, can I smoke when I pray?''
And he was severely reprimanded.
A second novice asked the prior: ``Father, can I pray when I smoke?''
And he was praised for his devotion.
\end{quote}

\noindent We see that \emph{pray \& smoke} does not elicit the same meanings as \emph{smoke \& pray}. In the same way, the perceived meaning of the joint question \emph{Are Clinton \& Gore honest?} is not exactly the same as that of the question \emph{Are Gore \& Clinton honest?} Accordingly, the perceived meaning of the answer \emph{Clinton is honest \& Gore is honest} is not the same as that of the answer \emph{Gore is honest \& Clinton is honest}, i.e, they do not correspond to the same state. This means that the two measurements $AB$ and $BA$, when interpreted as single measurements, their outcome states will be in general different, i.e., $AB$ and $BA$ will be described in the quantum formalism by two different non-commuting Hermitian operators. So, even though $AB$ and $BA$ are in practice executed as two-step processes, i.e., as processes during which an outcome state is created in a sequential way, one can wonder to which extent one is allowed to experimentally disentangle the sequence into two distinct measurements. In other words, in general, measurements $A$ and $B$ are to be considered entangled in the combinations $AB$ and $BA$ (for the notion of entangled measurements in cognition, see \cite{as2014}).

\section{A quantum equality\label{quantum-test}}

In this section, we consider measurements only having two (possibly degenerate) outcomes, and the following quantity \cite{WangBusemeyer2013,Niestegge2008}: 
\begin{equation}
q = p(a_1b_1)-p(b_1a_1)+p(a_2b_2) -p(b_2a_2).
\label{q}
\end{equation}
If we assume that probabilities have to be modeled as sequential processes, and that background information is also provided, which we also want it to be modeled at the level of the collective participant, 
then according to (\ref{option1}) for the first term of (\ref{q}) we can write: 
\begin{equation}
p(a_1b_1)= \mu(a_1,e^{\rm ind}_A,e^{\rm det}_A p_{\rm in})\mu(b_1,e^{\rm ind}_B,e^{\rm det}_B a_1),
\label{first-term}
\end{equation}
and similarly for the other terms. Let us model the above using the standard quantum formalism. The initial state $p_{\rm in}$ is then described by a ket $|\psi_{\rm in}\rangle\in {\cal H}$, where ${\cal H}$ denotes a Hilbert space of arbitrary dimension. Also, the two indeterministic contexts $e^{\rm ind}_A$ and $e^{\rm ind}_B$ are necessarily the same and their action is described by the Born rule and corresponding L\"uders-von Neumann projection formula. Finally, the two deterministic contexts $e^{\rm det}_A$ and $e^{\rm det}_B$ can be associated with two unitary operators, which we denote $U_A$ and $U_B$, respectively. For the first factor on the r.h.s. of (\ref{first-term}), we can write: $\mu(a_1,e^{\rm ind}_A,e^{\rm det}_A p_{\rm in})=\|P_AU_A|\psi_{\rm in}\rangle\|^2$, where $P_A$ denotes the orthogonal projection operator onto the subspace associated with the outcome-state $a_1$ of observable $A$, described by the vector $|a_1\rangle ={P_AU_A|\psi_{\rm in}\rangle \over \|P_AU_A|\psi_{\rm in}\rangle\|}$ (assuming that $P_AU_A|\psi_{\rm in}\rangle\neq 0$). Therefore, the second factor on the r.h.s. of (\ref{first-term}) can be written: $\mu(b_1,e^{\rm ind}_B,e^{\rm det}_B a_1)=\|P_BU_B|a_1\rangle\|^2= {\|P_BU_BP_AU_A|\psi_{\rm in}\rangle\|^2 \over \|P_AU_A|\psi_{\rm in}\rangle\|^2}$, where $P_B$ denotes the projection onto the subspace associated with the outcome-state $b_1$ of observable $B$. Multiplying these two factors, we thus find: 
\begin{equation}
p(a_1b_1)=\|P_BU_BP_AU_A|\psi_{\rm in}\rangle\|^2 =\langle \psi_{\rm in}|U_A^\dagger P_A U_B^\dagger P_BU_BP_AU_A|\psi_{\rm in}\rangle.
\label{first-term2}
\end{equation}
Proceeding in the same way with the other terms in (\ref{q}), one obtains that $q=\langle \psi_{\rm in}|Q|\psi_{\rm in}\rangle$, with the self-adjoint operator $Q$ given by:
\begin{equation}
Q = U_A^\dagger P_AP'_BP_AU_A - U_B^\dagger P_BP'_AP_BU_B + U_A^\dagger \bar{P}_A\bar{P}'_B\bar{P}_AU_A - U_B^\dagger \bar{P}_B\bar{P}'_A\bar{P}_BU_B,
\label{QQ}
\end{equation}
where we have defined the orthogonal projectors: $P'_A\equiv U_A^\dagger P_AU_A$, $P'_B\equiv U_B^\dagger P_BU_B$, $\bar{P}_A=\mathbb{I}-P_A$, $\bar{P}_B=\mathbb{I}-P_B$, $\bar{P}'_B=\mathbb{I}-P'_B$ and $\bar{P}'_A=\mathbb{I}-P'_A$. We have: 
\begin{eqnarray}
\lefteqn{U_A^\dagger \bar{P}_A\bar{P}'_B\bar{P}_AU_A =U_A^\dagger (\mathbb{I}-P_A)\bar{P}'_B\bar{P}_AU_A = U_A^\dagger \bar{P}'_B\bar{P}_AU_A - U_A^\dagger P_A\bar{P}'_B\bar{P}_AU_A}\nonumber\\
&&= U_A^\dagger \bar{P}'_B(\mathbb{I}-P_A)U_A - U_A^\dagger P_A\bar{P}'_B(\mathbb{I}-P_A)U_A\nonumber\\
&&=U_A^\dagger \bar{P}'_BU_A -U_A^\dagger \bar{P}'_BP_AU_A - U_A^\dagger P_A(\mathbb{I}-P'_B)U_A +U_A^\dagger P_A(\mathbb{I}-P'_B)P_AU_A\nonumber\\
&&=U_A^\dagger (\mathbb{I}-P'_B)U_A -U_A^\dagger (\mathbb{I}-P'_B)P_AU_A - U_A^\dagger P_A(\mathbb{I}-P'_B)U_A +U_A^\dagger P_A(\mathbb{I}-P'_B)P_AU_A\nonumber\\
&&=\mathbb{I} -U_A^\dagger P'_BU_A - U_A^\dagger P_AU_A + U_A^\dagger P'_BP_AU_A -U_A^\dagger P_AU_A + U_A^\dagger P_AP'_BU_A + U_A^\dagger P_AP_AU_A -U_A^\dagger P_AP'_BP_AU_A\nonumber\\
&&= \mathbb{I} -U_A^\dagger P'_BU_A - U_A^\dagger P_AU_A + U_A^\dagger P'_BP_AU_A + U_A^\dagger P_AP'_BU_A -U_A^\dagger P_AP'_BP_AU_A.\nonumber
\end{eqnarray}
Therefore: 
\begin{eqnarray}
U_A^\dagger P_AP'_BP_AU_A + U_A^\dagger \bar{P}_A\bar{P}'_B\bar{P}_AU_A &=&\mathbb{I} -U_A^\dagger P'_BU_A - U_A^\dagger P_AU_A + U_A^\dagger P'_BP_AU_A + U_A^\dagger P_AP'_BU_A\nonumber\\
&=&\mathbb{I} -U_A^\dagger P'_BU_A - P'_A + U_A^\dagger P'_BU_A P'_A + P'_AU_A^\dagger P'_BU_A.
\end{eqnarray}
Similarly, exchanging the roles of $A$ and $B$, we obtain:
\begin{eqnarray}
U_B^\dagger P_BP'_AP_BU_B + U_B^\dagger \bar{P}_B\bar{P}'_A\bar{P}_BU_B 
&=& \mathbb{I} -U_B^\dagger P'_AU_B - U_B^\dagger P_BU_B + U_B^\dagger P'_AP_BU_B + U_B^\dagger P_BP'_AU_B\nonumber\\
&=& =\mathbb{I} -U_B^\dagger P'_AU_B - P'_B + U_B^\dagger P'_AU_B P'_B + P'_BU_B^\dagger P'_AU_B.
\end{eqnarray}
The difference of the above two expressions then gives \cite{asdb2017b}: 
\begin{equation}
Q=(P'_B - U_A^\dagger P'_BU_A) +(U_B^\dagger P'_AU_B -P'_A) + (U_A^\dagger P'_BU_A P'_A - P'_BU_B^\dagger P'_AU_B) +\, (P'_AU_A^\dagger P'_BU_A - U_B^\dagger P'_AU_B P'_B).
\label{QQ2}
\end{equation}

We see that in general $Q\neq 0$, so that the average $q=\langle \psi_{\rm in}|Q|\psi_{\rm in}\rangle$ can in principle take any value inside the interval $[-1,1]$. However, if we consider that the modeling should not explicitly include deterministic contexts, or that they would be trivial, then we can set $U_A=U_B=\mathbb{I}$ in (\ref{QQ2}), and we clearly obtain $Q=0$, so that for every initial state we have the remarkable equality $q=0$. Clearly, if experimental data obey the  latter, they possibly (although not necessarily) have a pure quantum structure, whereas if the $q=0$ equality is disobeyed, one has to use beyond-quantum probabilistic models to fit the data.

\section{Testing the collective participant\label{violating-quantum-equalities}}

In this section, we submit the collective participant 
to the $q$-test derived in the previous section. More precisely, we show by means of a simple example that although the collective participant  
can behave in a pure quantum way on single measurements, when sequential measurements are performed at the individual level, the modeling of the obtained statistics of outcomes at the collective level will generally be non-quantum, as it will violate the $q=0$ equality. To do so, we place ourselves in the simplest possible situation: that of an experiment only using two participants. We also assume that they both select outcomes in a way that is independent of the measurement considered, characterized by the probability distributions $\rho_1(y)$ and $\rho_2(y)$, which are such that $\rho_1(y)+\rho_2(y)=1$. This means that the average probability (\ref{average}) exactly corresponds to the Born quantum probability, characterized by the uniform probability density $\rho_{\rm u}(y)={1\over 2}$. In other words, The collective participant, 
describing the average behavior of these two individuals, is associated with a pure quantum context $e_{\rho_{\rm u}}$, described by a uniform $\rho_{\rm u}$-way of selecting an outcome. 

In the following, we also consider for simplicity that no processes of information supply are to be considered. Eqs. (\ref{average4}) then become:
\begin{equation}
p(a_jb_k)= {1\over 2}\sum_{i=1}^2 \mu(a_j,e_{\rho_i},p_{\rm in})\mu(b_k,e_{\rho_i}, a_j), \quad p(b_ka_j)= {1\over 2}\sum_{i=1}^2 \mu(b_k,e_{\rho_i},p_{\rm in})\mu(a_j,e_{\rho_i}, b_k),
\label{equation}
\end{equation}
so that (\ref{q}) becomes: 
\begin{equation}
\begin{aligned}
2q&=2 [p(a_1b_1)-p(b_1a_1)+p(a_2b_2) -p(b_2a_2)]\\
&+\, \mu(a_1,e_{\rho_1},p_{\rm in})\mu(b_1,e_{\rho_1}, a_1)+\mu(a_1,e_{\rho_2},p_{\rm in})\mu(b_1,e_{\rho_2}, a_1)\\
&-\, \mu(b_1,e_{\rho_1},p_{\rm in})\mu(a_1,e_{\rho_1}, b_1) -\mu(b_1,e_{\rho_2},p_{\rm in})\mu(a_1,e_{\rho_2}, b_1)\\
&+\, \mu(a_2,e_{\rho_1},p_{\rm in})\mu(b_2,e_{\rho_1}, a_2)+\mu(a_2,e_{\rho_2},p_{\rm in})\mu(b_2,e_{\rho_2}, a_2)\\
&-\, \mu(b_2,e_{\rho_1},p_{\rm in})\mu(a_2,e_{\rho_1}, b_2) -\mu(b_2,e_{\rho_2},p_{\rm in})\mu(a_2,e_{\rho_2}, b_2).
\label{q-applied}
\end{aligned}
\end{equation}
Using $\mu(a_2,e_{\rho_1},p_{\rm in})=1- \mu(a_1,e_{\rho_1},p_{\rm in})$ and $\mu(b_2,e_{\rho_1},p_{\rm in})=1- \mu(b_1,e_{\rho_1},p_{\rm in})$, we obtain: 
\begin{equation}
\begin{aligned}
2q = &+\, \mu(a_1,e_{\rho_1},p_{\rm in})[\mu(b_1,e_{\rho_1}, a_1)-\mu(b_2,e_{\rho_1}, a_2)]+\mu(a_1,e_{\rho_2},p_{\rm in})[\mu(b_1,e_{\rho_2}, a_1)-\mu(b_2,e_{\rho_2}, a_2)]\\
& -\, \mu(b_1,e_{\rho_1},p_{\rm in})[\mu(a_1,e_{\rho_1}, b_1)- \mu(a_2,e_{\rho_1}, b_2)] - \mu(b_1,e_{\rho_2},p_{\rm in})[\mu(a_1,e_{\rho_2}, b_1)- \mu(a_2,e_{\rho_2}, b_2)]\\
& +\,\mu(b_2,e_{\rho_1}, a_2)- \mu(a_2,e_{\rho_1}, b_2) +\mu(b_2,e_{\rho_2}, a_2) -\mu(a_2,e_{\rho_2}, b_2).
\label{q-applied-2}
\end{aligned}
\end{equation}
We can observe that the third line of (\ref{q-applied-2}) is zero if $\mu(b_2,e_{\rho_1}, a_2) = \mu(a_2,e_{\rho_1}, b_2)$ and $\mu(b_2,e_{\rho_2}, a_2) =\mu(a_2,e_{\rho_2}, b_2)$, which will generally be the case for quantum transition probabilities. In the more general situation where measurements are not characterized by a globally uniform probability density $\rho_{\rm u}$, this can still be the case if individuals select outcomes in the same way in $A$ and $B$ measurements, which is what we have also previously assumed, for simplicity. Concerning the first line of (\ref{q-applied-2}), we observe it is zero if $\mu(b_1,e_{\rho_1}, a_1)=\mu(b_2,e_{\rho_1}, a_2)$ and $\mu(b_1,e_{\rho_2}, a_1)=\mu(b_2,e_{\rho_2}, a_2)$. For this to be so, we need $\rho_1$ and $\rho_2$ to be symmetric with respect to the origin of the Bloch sphere, i.e., $\rho_1(y)=\rho_1(-y)$ and $\rho_2(y)=\rho_2(-y)$, which of course will not be true in general. Similarly, the second line of (\ref{q-applied-2}) is zero if $\mu(a_1,e_{\rho_1}, b_1)= \mu(a_2,e_{\rho_1}, b_2)$ and $\mu(a_1,e_{\rho_2}, b_1)= \mu(a_2,e_{\rho_2}, b_2)$, which again requires $\rho_1$ and $\rho_2$ to be symmetric. 

So, if $\rho_1$ and $\rho_2$ are symmetric, and if we assume that the two individuals select outcomes in the same way (although not necessarily as per the Born rule) in the two measurements, then the $q=0$ equality is obeyed. But in general situations this will not be the case, so that the $q$-test will not be passed. As a simple example, consider $\rho_1(y)=\chi_{[-1,0]}(y)$ and $\rho_2(y)=\chi_{[0,1]}(y)$, where $\chi_{I}(y)$ denotes the characteristic function of interval $I$. If $\alpha$ denotes the the angle between the $\rho_1$-elastic and the $\rho_2$-elastic, i.e., ${\bf a}_1\cdot {\bf b}_1 = \cos\alpha$, and assuming for simplicity that $p_{\rm in} =a_1$, we have: $\mu(a_1,e_{\rho_1},a_1)=\mu(a_1,e_{\rho_2},a_1)=\mu(b_2,e_{\rho_2}, a_2)=\mu(a_2,e_{\rho_2}, b_2)=\mu(b_1,e_{\rho_1}, a_1)=\mu(a_1,e_{\rho_1}, b_1)=1$, $\mu(b_2,e_{\rho_1}, a_2)=\mu(a_2,e_{\rho_1}, b_2)=\mu(b_1,e_{\rho_2}, a_1) =\mu(a_1,e_{\rho_2}, b_1)=\cos\alpha$. Inserting these values in (\ref{q-applied-2}), we obtain: $q= -{1\over 2}(1-\cos\alpha)^2$, which is clearly different from zero for $\alpha\neq 0$. 
%\begin{eqnarray} &2q = 1[1-\cos\alpha]+1[\cos\alpha-1]-1[1-\cos\alpha] - \cos\alpha[\cos\alpha-1]+\cos\alpha- \cos\alpha +1 -1\\ &=-1+\cos\alpha -\cos^2\alpha + \cos\alpha = -(1-2\cos\alpha+\cos^2\alpha)=-(1-\cos\alpha)^2 \end{eqnarray}

Note that in all known sequential measurements the $q=0$ equality is violated: in some of them only very weakly, in others quite strongly, showing that the underlying probability model is intrinsically non-Hilbertian. The measurements where the violation is stronger appear to be those where some background information is provided to the participants, before answering the questions \cite{asdb2017a,asdb2017b,bb2012,WangBusemeyer2013,WangEtal2014}. Since in this case $Q$, given by (\ref{QQ2}), is different from zero, apparently the above description in terms of pure Bornian sequential measurements seems to provide an interesting modeling of the measurements, so much so that it is even able to predict, with good approximation, the $q=0$ remarkable relation. However, it can be objected that what is really important is not if the $q$ value is small, because even a small value is, strictly speaking, a violation, but if such value goes to zero as the number $n$ of participants increases, which as far as we know is something that has not been studied yet. Note also, as we mentioned already, that even an exact obedience of the $q=0$ equality would be insufficient to deduce that the underlying probability model is purely quantum, as probabilities that are more general than the Born probabilities are also able to obey the $q=0$ identity; see \cite{asdb2017a} for a more specific analysis of this aspect.

\section{Conclusion}

In this article, we emphasized the importance of distinguishing between the individual level of the participants in a psychological experiment, and their collective level, which we have associated with a notion of collective participant. When the latter only describes single measurement situations, one can generally expect the standard quantum formalism to provide an effective modeling tool for the data. This because when averaging over all possible ways of selecting an outcome, one recovers the Born quantum rule, so that a measurement associated with a collective participant who is the expression of a sufficiently representative collection of individuals can be expected to be structurally close to a pure quantum measurement. 

The modeling becomes however much more involved when individuals perform more than a single measurement, in a sequential way. Then, one should not expect anymore the standard quantum formalism to be sufficient to model typical data. In the article, we have shown this by considering the simplest possible situation of a collective participant  formed by only two individuals, whose collective action is pure-quantum when only single (non-sequential) measurements are considered, but irreducibly beyond-quantum (non-Born) when at the individual level outcomes are selected in a sequential way, so that a more general mathematical framework is needed to model the obtained experimental probabilities \cite{asdb2017a,asdb2017b}.

In our analysis, we have also proposed a distinction between deterministic and indeterministic cognitive processes/contexts, as formalized in the decomposition (\ref{decomposition}) of the (possibly different) individual contexts. As far as this distinction is concerned, we observed that different modeling options are possible, putting the cognitive action more in the deterministic evolution of the initial state or in the indeterministic collapse of it, or in some combination of the two. Concerning the individual cognitive processes, it is however clear that the different possible ways of modeling them should correspond to objective components of the cognitive situations, possibly testable in well-designed experiments.

\end{document}